


\font\titlefont = cmr10 scaled\magstep 4
\font\sectionfont = cmr10
\font\littlefont = cmr5
\font\eightrm = cmr8

\def\ss{\scriptstyle}
\def\sss{\scriptscriptstyle}

\newcount\tcflag
\tcflag = 0  

\ifnum\tcflag = 0 \magnification = 1200 \fi  

\global\baselineskip = 1.2\baselineskip
\global\parskip = 4pt plus 0.3pt
\global\abovedisplayskip = 18pt plus3pt minus9pt
\global\belowdisplayskip = 18pt plus3pt minus9pt
\global\abovedisplayshortskip = 6pt plus3pt
\global\belowdisplayshortskip = 6pt plus3pt

\def\barsoff{\overfullrule=0pt}


\def\endignore{}
\def\ignore #1\endignore{}

\newcount\dflag
\dflag = 0


\def\monthname{\ifcase\month
\or January \or February \or March \or April \or May \or June%
\or July \or August \or September \or October \or November %
\or December
\fi}

\newcount\dummy
\newcount\minute  
\newcount\hour
\newcount\localtime
\newcount\localday
\localtime = \time
\localday = \day

\def\advanceclock#1#2{ 
\dummy = #1
\multiply\dummy by 60
\advance\dummy by #2
\advance\localtime by \dummy
\ifnum\localtime > 1440 
\advance\localtime by -1440
\advance\localday by 1
\fi}

\def\settime{{\dummy = \localtime%
\divide\dummy by 60%
\hour = \dummy
\minute = \localtime%
\multiply\dummy by 60%
\advance\minute by -\dummy
\ifnum\minute < 10
\xdef\spacer{0} 
\else \xdef\spacer{}
\fi %
\ifnum\hour < 12
\xdef\ampm{a.m.} 
\else
\xdef\ampm{p.m.} 
\advance\hour by -12 %
\fi %
\ifnum\hour = 0 \hour = 12 \fi
\xdef\timestring{\number\hour : \spacer \number\minute%
\thinspace \ampm}}}



\def\endtitle{}
\def\title#1\endtitle{\vskip.5in\titlefont
\global\baselineskip = 2\baselineskip
#1\vskip.4in
\baselineskip = 0.5\baselineskip\rm}

\def\endauthors{}
\def\authors#1\endauthors{#1}

\def\endabstract{}
\def\abstract#1\endabstract{\vskip .3in%
\centerline{\sectionfont\bf Abstract}%
\vskip .1in
\noindent#1}

\def\nopageonenumber{\footline={\ifnum\pageno<2\hfil\else
\hss\tenrm\folio\hss\fi}}  

\newcount\nsection
\newcount\nsubsection

\def\section#1{\global\advance\nsection by 1
\nsubsection=0
\bigskip\noindent\centerline{\sectionfont \bf \number\nsection.\ #1}
\bigskip\rm\nobreak}

\def\subsection#1{\global\advance\nsubsection by 1
\bigskip\noindent\sectionfont \sl \number\nsection.\number\nsubsection)\
#1\bigskip\rm\nobreak}


\def\appendix#1#2{\bigskip\noindent%
\centerline{\sectionfont \bf Appendix #1.\ #2}
\bigskip\rm\nobreak}


\newcount\nref
\global\nref = 1

\def\therefs{}


\def\ref#1#2{\xdef #1{[\number\nref]}
\ifnum\nref = 1\global\xdef\therefs{\item{[\number\nref]} #2\ }
\else
\global\xdef\oldrefs{\therefs}
\global\xdef\therefs{\oldrefs\vskip.1in\item{[\number\nref]} #2\ }%
\fi%
\global\advance\nref by 1
}

\def\listrefs{\vfill\eject\section{References}\therefs}


\newcount\nfoot
\global\nfoot = 1

\def\foot#1#2{\xdef #1{(\number\nfoot)}
\footnote{${}^{\number\nfoot}$}{\eightrm #2}
\global\advance\nfoot by 1
}


\newcount\nfig
\global\nfig = 1
\def\thefigs{} 

\def\figure#1#2{\xdef #1{(\number\nfig)}
\ifnum\nfig = 1\global\xdef\thefigs{\item{(\number\nfig)} #2\ }
\else
\global\xdef\oldfigs{\thefigs}
\global\xdef\thefigs{\oldfigs\vskip.1in\item{(\number\nfig)} #2\ }%
\fi%
\global\advance\nfig by 1 } 

\def\figurecaptions{\vfill\eject\section{Figure Captions}\thefigs}

\def\fig#1{\xdef #1{(\number\nfig)}
\global\advance\nfig by 1 } 


\newcount\cflag
\newcount\nequation
\global\nequation = 1
\def\eqlabel{(1)}

\def\nexteqno{\ifnum\cflag = 0
\global\advance\nequation by 1
\fi
\global\cflag = 0
\xdef\eqlabel{(\number\nequation)}}

\def\lasteqno{\global\advance\nequation by -1
\xdef\eqlabel{(\number\nequation)}}

\def\label#1{\xdef #1{(\number\nequation)}
\ifnum\dflag = 1
{\escapechar = -1
\xdef\draftname{\littlefont\string#1}}
\fi}

\def\clabel#1#2{\xdef\eqlabel{(\number\nequation #2)}
\global\cflag = 1
\xdef #1{\eqlabel}
\ifnum\dflag = 1
{\escapechar = -1
\xdef\draftname{\string#1}}
\fi}

\def\cclabel#1#2{\xdef\eqlabel{#2)}
\global\cflag = 1
\xdef #1{\eqlabel}
\ifnum\dflag = 1
{\escapechar = -1
\xdef\draftname{\string#1}}
\fi}


\def\eeq{}

\def\eqnn #1\eeq{$$ #1 $$}

\def\eq #1\eeq{
\ifnum\dflag = 0
{\xdef\draftname{\ }}
\fi 
$$ #1
\eqno{\eqlabel \rlap{\ \draftname}} $$
\nexteqno}







\def\eqa #1\eeq{
\ifnum\dflag = 0
{\xdef\draftname{\ }}
\fi 
$$ \eqalignno{ #1 } $$
\global\cflag = 0}


\def\ie{{\it i.e.\/}}
\def\eg{{\it e.g.\/}}

\def\etal{{\it et.al.\/}}

\def\cf{{\it c.f.\/}}


\def\mpla#1#2#3{{\it Mod.\ Phys.\ Lett.} {\bf A#1}, (19#2) #3}

\def\npb#1#2#3{{\it Nucl.\ Phys.} {\bf B#1} (19#2) #3}
\def\plb#1#2#3{{\it Phys.\ Lett.} {\bf #1B} (19#2) #3}

\def\prd#1#2#3{{\it Phys.\ Rev.} {\bf D#1} (19#2) #3}
\def\pr#1#2#3{{\it Phys.\ Rev.} {\bf #1} (19#2) #3}

\def\prl#1#2#3{{\it Phys.\ Rev.\ Lett.} {\bf #1} (19#2) #3}


\global\nulldelimiterspace = 0pt



\def\frac#1#2{{{#1} \over {#2}}\,}  
\def\hf{{1\over 2}}



\def\Dsl{\hbox{/\kern-.6700em\it D}} 
\def\dsl{\hbox{/\kern-.5300em$\partial$}}
\def\pxpsl{\hbox{/\kern-.5600em$p$}}
\def\ssl{\hbox{/\kern-.5300em$s$}}
\def\epssl{\hbox{/\kern-.5100em$\epsilon$}}
\def\delsl{\hbox{/\kern-.6300em$\nabla$}}
\def\lxpsl{\hbox{/\kern-.4300em$l$}}
\def\elxpsl{\hbox{/\kern-.4500em$\ell$}}
\def\kxpsl{\hbox{/\kern-.5100em$k$}}
\def\qxpsl{\hbox{/\kern-.5000em$q$}}
\def\sla#1{\raise.15ex\hbox{$/$}\kern-.57em #1}



\def\roughly#1{\mathrel{\raise.3ex\hbox{$#1$
\kern-.75em\lower1ex\hbox{$\sim$}}}}

\def\tw#1{\tilde{#1}}
\def\ol#1{\overline{#1}}





\def\Scl{{\cal L}}


\def\ssa{{\sss A}}

\def\ssd{{\sss D}}
\def\sse{{\sss E}}

\def\ssl{{\sss L}}
\def\ssm{{\sss M}}


\def\det{\mathop{\rm det}}






\nopageonenumber
\baselineskip = 16pt
\barsoff


\def\sltwor{$SL(2,\IR)$}
\def\sltwoz{$SL(2,\IZ)$}
\def\ooneone{$O(1,1)$}
\def\abar{\overline{a}}

\def\ap{\alpha'}

\def\Fbar{\overline{F}}
\def\Gbar{\overline{G}}
\def\Hbar{\overline{H}}

\def\gbar{\overline{g}}

\def\phibar{\overline{\phi}}
\def\Abar{\overline{A}}
\def\Bbar{\overline{B}}

\def\abar{\overline{a}}

\def\Nbar{\overline{N}}
\def\Qbar{\overline{Q}}

\def\GN{G_{\scriptscriptstyle N}}
\def\bk{\item{}}
\def\IR{\relax{\rm I\kern-.18em R}}
\def\cam{{\cal M}}
\def\camb{\overline{\cal M}}
\def\cak{{\cal K}}
\def\IR{\relax{\rm I\kern-.18em R}}
\font\cmss=cmss10 \font\cmsss=cmss10 at 7pt
\def\IZ{\relax\ifmmode\mathchoice
{\hbox{\cmss Z\kern-.4em Z}}{\hbox{\cmss Z\kern-.4em Z}}
{\lower.9pt\hbox{\cmsss Z\kern-.4em Z}}
{\lower1.2pt\hbox{\cmsss Z\kern-.4em Z}}\else{\cmss Z\kern-.4em Z}\fi}


\line{hep-th/9410142 \hfil McGill-94/47, NEIP-94-011}

\title
\centerline{On Spherically Symmetric String Solutions}
\centerline{in Four Dimensions}
\endtitle

\authors
\centerline{C.P. Burgess,${}^a$ R.C. Myers,${}^a$ and F. Quevedo${}^b$}
\vskip .15in
\centerline{\it ${}^a$ Physics Department, McGill University}
\centerline{\it 3600 University St., Montr\'eal, Qu\'ebec, Canada, H3A 2T8.}
\vskip .1in
\centerline{\it ${}^b$ Institut de Physique, Universit\'e de Neuch\^atel}
\centerline{\it CH-2000 Neuch\^atel, Switzerland.}
\endauthors

\vskip .2in
\abstract
We reconsider here the problem of finding the general 4D spherically
symmetric, asymptotically flat and time-independent solutions
to the lowest-order string equations in the $\ap$ expansion.  Our
construction includes earlier work, but differs from it in three ways.
(1) We work with general background metric, dilaton, axion
and $U(1)$ gauge fields. (2) Much of the original solutions were required
to be nonsingular at the apparent
horizon, motivated by an interest in finding string corrections to
black hole spacetimes. We relax this condition throughout, motivated
by the  realization that string theory has a less restrictive
notion of what constitutes a singular field configuration
than do point particle theories. (3) We
can construct the general solution from a particularly simple one,
by generating it from successive applications of the {\it noncommuting}
\sltwor\  and \ooneone\  symmetries of the low-energy string equations
containing $S$ and target--space dualities respectively.
This allows its construction using relatively simple, purely algebraic,
techniques. The general solution is determined by the asymptotic behaviour
 of the various
fields: \ie\ by the mass, dilaton charge, axion charge, electric charge,
magnetic charge, and Taub-NUT parameter.
\endabstract


\vfill\eject

\section{Introduction}

Understanding the ultimate fate of a runaway gravitational collapse has been
a longstanding problem ever since its discovery as a prediction of General
Relativity (GR) many years ago. String theory is perhaps the only
presently-known
theory which has pretentions to describe physics at the Planck scale, and so
potentially to provide some insight into this problem. The challenge has been
to reliably compute string behaviour in the presence of very strong
gravitational fields.

Since gravitational collapse is a classical phenomenon, the simplest approach
is to investigate the corresponding solutions to the classical string
equations.
Classical string theory modifies classical GR in at least two ways. Firstly,
the string field equations for the metric only reproduce those of GR in the
limit of weak curvatures in comparison to the natural string scale (typically
parameterized by $\ap$).\foot\units{We use fundamental units, for which
$\ss \hbar = c = 1$, and so $\ss \ap$ is of order the Planck length squared.}
In situations of strong curvature, higher derivative terms in the
effective field equations will become significant.
Secondly, string theory introduces additional light degrees of freedom,
beyond the metric, which typically cannot vanish in nontrivial solutions
to the full string equations.

\ref\progress{For a recent review see, \eg\ A.A. Tseytlin, preprint
Imperial/TP/93-94/46, hep-th/9407099.}
\ref\duality{For a recent review see, \eg\ A. Giveon, M. Porrati and E.
Rabinovici,
preprint RI-1-94, hep-th/9401139.}
\ref\twodbh{E. Witten, \prd{44}{91}{314}}
\ref\moretwodbh{R. Dijkgraaf, E. Verlinde and
H. Verlinde, \npb{371}{92}{269};\bk A. Giveon \mpla{6}{91}{2843}.}
\ref\bpb{J. Horne and G. Horowitz, \npb{368}{92}{444};\bk
J. Horne, G. Horowitz and A. Steif, \prl{68}{92}{568}.}
\ref\paulandme{P. Ginsparg and F. Quevedo, \npb{385}{92}{527}.}

In fact, there has been real progress in the understanding of the properties of
strings in the presence of more complicated background fields over the last ten
years. This progress has included ($i$) the construction \progress\ of
strongly-curved field configurations which are known to be solutions to the
full string equations; ($ii$) the discovery of `duality' transformations
\duality,
which relate superficially very different, but often actually physically
identical, string configurations; and ($iii$) the application of these two
tools to
the detailed exploration of black-hole configurations in two spacetime
dimensions \twodbh, \moretwodbh, and to black-$p$-brane configurations
in higher dimensions \bpb, \paulandme, which are known as exact
conformal field theories.

One of the surprising features to emerge from these developments has been the
realization that string theory may be quite forgiving in its notion of what
constitutes a physically unacceptable singularity. What appears to be a
malignantly singular field configuration from the point of view
of point-particle theory, can be completely benign as a background
for string propagation. The duality transformation constructed using a
rotation symmetry of flat space furnishes a particularly striking example
of this, since it produces a curved manifold with a
curvature singularity at the rotation axis. Similarly, for two-dimensional
black holes the nonsingular horizon is mapped by duality into
the curvature singularity at $r=0$, and for three-dimensional black strings
the singularity is mapped to a regular surface in the asymptotically-flat
region.

\ref\strbhone{C.G. Callan, R.C. Myers and M.J. Perry, \npb{311}{89}{673}.}
\ref\strbhtoo{R.C. Myers, \npb{289}{87}{701}.}

Taking seriously this broader perspective concerning the potential
acceptability
of singular field configurations has some immediate implications for the study
of classical string configurations. In particular, the point of departure for
studies of string propagation through complicated
backgrounds has usually been the construction of solutions
to the approximate string equations, to lowest order in $\ap$.
Interestingly, the string corrections are typically
singular at the apparent horizon of the lowest-order black hole
solutions of GR, although these singularities can be avoided by making
an appropriate choice for the boundary conditions for the new
fields, such as the dilaton. This observation led early workers \strbhone,
\strbhtoo\ to discard those solutions for which this adjustment
was not made.

\ref\moresolns{M.J. Bowick, S.B. Giddings, J.A. Harvey, G.T. Horowitz,
and A. Strominger, \prl{61}{88}{2823}.}
\ref\dark{For a review, see: G.T. Horowitz,  ``The Dark Side of String
Theory: Black Holes and Black Strings,'' in Proceedings of Trieste 1992,
hep-th/9210119}
\ref\gibbons{G.W. Gibbons and K. Maeda, \npb{298}{88}{741};\bk
D. Garfinkle, G. Horowitz and A. Strominger, \prd{43}{91}{3140}.}
\ref\wilczek{A. Shapere, S. Trivedi and F. Wilczek, \mpla{6}{91}{2677}.}
\ref\kall{R. Kallosh and T. Ort\'{\i}n \prd{48}{93}{742}.}
\ref\italia{A.G. Agnese and M. La Camera, \prd{49}{94}{2126}.}

The purpose of the present paper is to re-examine the solutions to the
low-energy string equations in four (and higher) dimensions.
Keeping in mind the observation that, in string theory, curvature
singularities need not be all that they seem, we construct the general
time-independent, spherically-symmetric and asymptotically flat
solution to the lowest-order string equations, and do {\it not} exclude
those configurations in which singularities are not hidden by an event
horizon. The nontrivial fields which we will consider are the metric, the
dilaton, the Kalb-Ramond field and an abelian vector potential.
The most general solutions would then be characterized by
five independent parameters corresponding to the configuration's
mass, dilaton charge, axion charge, electric charge, and magnetic charge.
Our construction will also naturally introduce a sixth parameter,
namely the Taub-NUT parameter.
All but the mass vanish in the usual Schwarzschild solution. (We do not
consider nonvanishing topological charges such as the `axion hair'
considered in Ref.~\moresolns.) The final six-parameter family includes, but
extends, many of the solutions that have been considered
heretofore \dark, \gibbons, \wilczek, \kall, \italia.

Rather than facing the daunting task of explicitly constructing the
solutions to the relevant coupled nonlinear PDE's, we instead construct
these solutions by exploiting some of the extraordinary symmetries of the
low-energy string equations. In particular, starting with the general
spherically symmetric, static and asymptotically flat solution to the
dilaton-metric system, we generate the others by successively applying
the {\it noncommuting} \sltwor\  and \ooneone\  symmetries of the
low-energy string equations containing $S$ duality and target
space duality respectively. This has the labour-saving advantage of
only requiring algebraic techniques.

\ref\wald{R.M. Wald, {\it General Relativity}, (1984) (University of
Chicago Press).}

The paper is organized as follows. In the next section we display our starting
two-parameter family of static, spherically-symmetric dilaton-metric
configurations in four dimensions. We follow this, in section 3,
by the extension of these results to the more general solution of the
 metric--dilaton--axion system, which is the generic case for the
closed bosonic string. Starting from the solutions of section 2, we
generate solutions with nonvanishing axion field --- \ie the antisymmetric
tensor field, $B_{\mu\nu}$ --- by performing an \sltwor\
transformation which is a symmetry of the low-energy field equations.
Applying a target-space duality
transformation to this result then produces
new solutions with nonvanishing Taub--NUT parameter but zero axion field.
These new solutions differ from our original ansatz in that they are
{\it stationary}, as opposed to being static \wald. Further, they are
only spherically symmetric in the generalized sense of being invariant
under rotations that are combined with a simultaneous position-dependent
time translation. A further \sltwor\  transformation then
generates a more general class of solutions with both a nonvanishing
Taub--NUT parameter and a nonzero axion field. This underlines
the fact that these two symmetries --- standard duality and \sltwor\
invariance --- {\it do not} commute.
Performing further duality transformations
to this general solution does not yield any new backgrounds.
In section 4, we extend this procedure to also include a nonzero
electromagnetic gauge potential.  We do so by using successive
applications of the continuous \ooneone\  symmetry (which contains
ordinary duality as a special case) together with the \sltwor\  symmetry.
We obtain in this way two more parameters in our family of solutions,
which can be identified with their electric and magnetic
charges. These results are summarized in our concluding section.
We include a (partial) generalization of these solutions to
higher dimensions as an Appendix.

\section{Spherically Symmetric Dilaton--Metric Solutions}

\ref\lestrings{See, \eg\ M. Green, J. Schwarz and E. Witten
{\it Superstring Theory I} (1987), (Cambridge University Press).}
The massless bosonic fields which always (in string
perturbation theory) appear in the spectrum of a generic string theory consist
of the metric, $G_{\mu \nu}$, the dilaton, $\phi$, an antisymmetric
Kalb-Ramond field, $B_{\mu \nu}$. In heterotic strings these
can also accompanied by one or more gauge potentials, $A_\mu$.
The Lagrangian density which governs these fields at low
energies is given by \lestrings:
\label\leaction
\eq
\Scl = {1 \over 8 \pi} \; \left( {1 \over \ap}\right)^{(d-2)/2} \; \sqrt{- G}
\; e^\phi
\left[ R(G) + (\nabla \phi)^2 - {1 \over 12} \, H^{\mu\nu\lambda}
H_{\mu\nu\lambda} - {1\over 8} \, F^{\mu\nu}F_{\mu\nu} \right] + \cdots \, ,
\eeq
where $H = dB + \hbox{(Chern-Simons terms)}$ and $F = dA$ are,
respectively,
the field strengths for the Kalb-Ramond and electromagnetic fields,
while $R(G)$ is the Ricci scalar for the so-called `sigma-model' metric,
$G_{\mu\nu}$, and $\sqrt{-G} = \sqrt{-\det G_{\mu\nu}}$. The ellipses
in eq.~\leaction\ represent terms which involve other massless fields
and/or terms involving more derivatives that arise at higher orders in
the $\ap$ expansion. We do not consider a cosmological constant in
\leaction\ since we assume that the solutions we
find are complemented by a corresponding conformal field theory
(such as a toroidal or Calabi--Yau compactification)
that saturates the central charge
to produce a full solution with
conformal invariance on the world-sheet.

It is sometimes useful to rescale the sigma-model
metric in order to ensure that the coefficient of the scalar curvature
is independent of $\phi$. In $d$ spacetime
dimensions this is accomplished by transforming to the `Einstein' metric,
\label\esmrelation
\eq  g_{\mu\nu} \equiv  e^{2 \phi /(d-2)} \, G_{\mu\nu}.
\eeq
In the following, we will denote the line element for the Einstein
metric by $ds^2$, while $dS^2$ will be reserved for that of the sigma-model
metric.

In the present section we set the antisymmetric tensor, $B_{\mu\nu}$, and
the gauge potential, $A_\mu$, to zero and solve for the general
dilaton--metric configuration. (We use the resulting solution to generate
more general axion- and gauge-potential-dependent field configurations
in the next two sections.) The relevant equations of motion then are
\label\eqofmotion
\eq \eqalign{
R_{\mu\nu}(g) &= {1\over d-2} \nabla_\mu \phi \nabla_\nu \phi \cr
\nabla^2 \phi &=0 \cr}
\eeq

\subsection{Lowest-Order Four-Dimensional Solutions}

We now turn to the solutions to these leading-order
low-energy string equations. We specialize to field configurations which are
explicitly static, spherically symmetric and asymptotically flat.
That is, we take:
\label\sssconfig
\eq
\phi = \phi(r),
\eeq
in coordinates for which the Einstein metric (in $d=4$ dimensions)
takes the form:
\label\ansatz
\eq
ds^2 = - f(r)dt^2 + {dr^2\over f(r)} + h^2(r) ( d\theta^2 + \sin^2 \theta \,
d\varphi^2 ).
\eeq

\ref\fifties{H. Buchdahl, \pr{115}{59}{1325};\bk
A.I. Janis, D.C. Robinson and J. Winicour, \pr{186}{69}{1729}.}

At large radius, asymptotic flatness implies that $f$ approaches unity
while $h$ approaches $r$.
In the same limit the dilaton asymptotically approaches a
constant, $\phi_0$. It is convenient in what follows if we absorb this
constant into the definition of Newton's constant, $\GN$. For instance, for
$d=4$, comparison of the action of eq.~\leaction\ with its standard form
gives $\GN = \hf  \, e^{-\phi_0} \, \ap$. With this choice in mind we can
choose $\phi \to 0$ at infinity, in which case the asymptotically flat
solutions to these equations become \fifties
\label\answer
\eq \eqalign{
f &= \left( 1 - {\ell\over r} \right)^\delta \cr
h^2 &= r^2 \left( 1 - {\ell\over r} \right)^{1-\delta} \cr
e^\phi &= \left( 1 - {\ell\over r} \right)^\gamma \cr}
\eeq
where $\ell$, $\delta$ and $\gamma$ are arbitrary
constants, subject to the one condition $\delta^2 + \gamma^2 = 1$.
(In the following, we will also assume that $\ell>0$ for simplicity.)
The choice $(\delta,\gamma)=(1,0)$
yields the standard Schwarzschild solution
with $\ell$ related to the black hole mass, $M$, according to $\ell = 2 \GN M$.
Up to a coordinate transformation, $(\delta,\gamma)=(-1,0)$ also corresponds
to a Schwarzschild black hole, albeit with $\ell = - 2 \GN M$.

\ref\ADM{R. Arnowitt, S. Deser and C.W. Misner, in {\it Gravitation:
An Introduction to Current Research}, ed. by L. Witten (New York, Wiley,
1962).}
\ref\weinberg{See, \eg\ S.~Weinberg, {\it Gravitation and Cosmology:
Principles and Applications of the General Theory of Relativity} (1982),
(New York: Wiley).}

Quite generally the two free parameters in this solution correspond to
the two quantities which label static, spherically-symmetric and
asymptotically flat dilaton--metric configurations: the mass, $M$,
and dilaton charge, $Q_\ssd$. We define the mass to be the conserved
(ADM) energy \ADM, which emerges in a calculation of the energy using a
gravitational stress-energy pseudo-tensor \weinberg,
using the Einstein metric. This is equivalent
to defining $2 \GN M$ to be the coefficient of $-(1/r)$ in the large-$r$
expansion of the function $f(r)$ which appears in the Einstein metric
ansatz, eq.~\ansatz. We similarly define the dilaton charge, $Q_\ssd$,
as the coefficient of $-(1/r)$ in the large-$r$ expansion of $\phi$.
For the solution of eq.~\answer\ we therefore have: $M = (\delta \ell/2 \GN)$
and $Q_\ssd = \gamma \ell$.

When $\gamma$ is not zero, the solutions contain a curvature singularity
at $r=\ell$, as can be seen from the equations of motion:
\label\singular
\eq \eqalign{
R(g) &= \hf \; (\nabla \phi)^2 \cr
&= {\gamma^2 \ell^2 \over 2 r^4} \left( 1 - {\ell \over r}
\right)^{\delta-2} . \cr}
\eeq
Notice that $|\delta|\le1$, and it is only at $\delta=\pm1$ that $\gamma=0$
and hence the above singularity vanishes. Of course, the latter solutions
are still singular at $r=0$ (even though $R=0$ there). Other solutions
with nonsingular horizons are also known when more string fields are present,
such as with gauge fields \gibbons, and antisymmetric Kalb-Ramond
fields \gibbons, \moresolns. We include those fields in the following sections.

\section{Axionic and Taub--NUT extensions}

\ref\oldsduality{E. Witten, \plb{155}{85}{151}; \bk
C.P. Burgess, A. Font and F. Quevedo, \npb{272}{86}{661}; \bk
S.P. Li, R. Peschanski and C. Savoy, \plb{194}{87}{226};
\plb{178}{86}{193}.}
\ref\slur{M. De Roo, \npb {255}{85}{515};\bk
A. Sen, \npb{404}{93}{109}.}

We next turn to the generalization of the
(leading order) four-dimensional solutions presented in section
2.1. We generate these new solutions by repeated applications
of \sltwor\  \oldsduality, \wilczek, \slur\ and discrete
target-space duality transformations. We restrict ourselves here to
spherically-symmetric, asymptotically flat, time independent
configurations involving the metric, dilaton and axion fields,
where the axion here represents the antisymmetric Kalb-Ramond
tensor, $B_{\mu\nu}$. We treat the case of background gauge fields
in the next section. (An overview of our final construction
is illustrated in the Figure.)

\figure\figone{The sequence of transformations with which
we generate the six-parameter family of solutions. The second column
shows the origins of each of the six parameters, $\ell$, $\delta$,
$\omega$, $\epsilon$, $x$ and $\rho$.}

\ref\taubnut{A.H. Taub, {\sl Ann. Math.} {\bf 53}
(1951) 472; E. Newman, L. Tamburino and T. Unti, {\sl J. Math. Phys.}
{\bf 4} (1963) 915.}
\ref\misner{C.W. Misner, `Taub-NUT space as a counterexample to almost
anything', in {\it Relativity Theory and Astrophysics I, Relativity and
Cosmology},
ed. by J. Ehlers, Lectures in Applied Mathematics, Vol. 8 (American
Mathematical
Society, 1967) 160.}

One of the interesting features of
our construction is that, besides introducing a nontrivial axion
background, repeated duality and \sltwor\  transformations also
force us to generalize our disposition on the character of the
solutions of interest. Up to this point we have taken
a {\it static} metric ansatz \wald,
for which the metric is independent of a time coordinate, $t$, and
curves  along which only $t$ varies are orthogonal to the hypersurfaces
of constant $t$. Repeated symmetry transformations, however, take
us to a metric which is only {\it stationary}, in that it is still $t$
independent although it is impossible to choose $t$ in a
`hypersurface orthogonal' way.  The stationary metrics that we find
are reminiscent of the `Taub-NUT' metric \taubnut, \misner.
Thus the solutions also become spherically symmetric only in the
modified sense that $SO(3)$ transformations are only symmetries
when they are compensated by an appropriate position-dependent
time translation.

In four dimensions, the antisymmetric tensor is dual to a pseudoscalar
field, $a(x)$, defined by
\label\axion
\eq
H_{\mu\nu\rho}=-e^{-2\phi}\epsilon_{\mu\nu\rho\kappa}\nabla^\kappa a
\eeq
where $H_{\mu\nu\rho}=\partial_\mu B_{\nu\rho}+\partial_\nu
B_{\rho\mu}+\partial_\rho B_{\mu\nu}+$ (Chern--Simons terms).
In eq.~\axion\ indices are raised and lowered with the Einstein metric
$g_{\mu\nu}$ and $\epsilon_{tr\theta\varphi}=\sqrt{-g}$.

\subsection{The Static Solution}

\ref\sduality{A. Font, L.E. Ib\'a\~nez, D. L\"ust and F. Quevedo,
\plb{249}{90}{35};\bk
S.J. Rey, \prd{43}{91}{526}.}

To find the axionic backgrounds we use the fact that the field equations
are invariant under a continuous \sltwor\  symmetry \oldsduality, \wilczek,
\slur\ acting on the complex field $S=a+i e^\phi$ as:
\label\sltr
\eq
S\longrightarrow {{\rm a}\, S + {\rm b}\over {\rm c}\, S +{\rm d}}
\eeq
where a, b, c and d are real number satisfying ${\rm ad-bc}=1$.
The Einstein metric is invariant under these transformations. The
discrete subgroup, \sltwoz\ , of these transformations which
contains strong--weak coupling duality (\ie, $S$--duality),
is also conjectured
to be a symmetry of the full quantum string theory, as well as the
leading order low-energy field equations \sduality.

Given eq.~\sltr, there is a three  parameter family
of backgrounds that can be generated by applying these
transformations
to any given four-dimensional solution. We start with
a vanishing axion configuration, $a = 0$, together with
the dilaton and (Einstein) metric backgrounds given
in \sssconfig, \ansatz\ and \answer. We repeat these
here for ease of reference:
\label\ansatzt
\eq
ds^2 = - f(r) \, dt^2 + {dr^2\over f(r)} + h^2(r) ( d\theta^2
+ \sin^2 \theta \, d\varphi^2 )
\eeq
with $\phi(r)$, $f(r)$ and $h(r)$ given as the following powers of
the quantity, $\Lambda(r) \equiv 1 - \ell/r$:
\label\answertt
\eq
 e^{\phi (r)} = \Lambda^\gamma, \qquad
f(r) = \Lambda^\delta, \qquad \hbox{and} \qquad
h^2(r) = r^2 \Lambda^{1-\delta} .
\eeq
Recall also that the parameters, $\delta$ and $\gamma$, are related by
$\delta^2+\gamma^2=1$.

Performing the \sltwor\  transformation of eq.~\sltr, we obtain new
dilaton and axion fields, $\hat\phi$ and $\hat a$:
\label\first
\eq
e^{\hat\phi}={\Lambda^\gamma\over {\rm c}^2\, \Lambda^{2\gamma}+{\rm d}^2}
\, \, ,
\qquad
\hat a={{\rm a c}\, \Lambda^{2\gamma} +{\rm b d}\over {\rm c}^2\,
\Lambda^{2\gamma}
+{\rm d}^2}
\eeq

Since the low-energy equations of motion are
invariant under constant shifts of the axion field, $a$,
we may choose the axion field to vanish asymptotically.
We also continue to impose the vanishing of the
dilaton at infinity. Only a one-parameter family
of the \sltwor\ transformations respects these conditions, however, given by
${\rm a=d=\sqrt{1-c^2},\,  b=-c}$). It is convenient to
use the ratio $\omega\equiv {\rm c/d}$ as the independent parameter,
in which case the dilaton and axion configurations of eq.~\first\ become:
\label\firstt
\eq
e^{\hat\phi} =\left(1+\omega^2\right){\Lambda^\gamma
\over\omega^2\Lambda^{2\gamma}+1} , \qquad
\hat a ={\omega\left(\Lambda^{2\gamma}-1\right) \over
\omega^2\Lambda^{2\gamma}+1} .
\eeq

It is also convenient to follow our practice for the dilaton, and so to
define the axion charge, $Q_\ssa$, as the coefficient of $-(1/r)$
in the large-$r$ expansion of the axion field, $a(r)$. With this definition
the dilaton and axion charges of the solutions of eq.~\firstt\ are
\label\charges
\eq
Q_{\ssd}= {1-\omega^2 \over 1+\omega^2}\; \gamma \, \ell \qquad
\hbox{ and} \qquad Q_\ssa={2\omega \, \gamma \, \ell\over 1+\omega^2},
\eeq
respectively.

The original antisymmetric field, $B_{\mu\nu}$, which corresponds,
using eq.~\axion, to the axion configuration, $\hat a$, has a particularly
simple expression in terms of the charge, $Q_\ssa$. It is:
\label\bfirst
\eq
\hat B_{\varphi t}=Q_\ssa \, \cos\theta\ ,
\eeq
which has no dependence on $r$.

The two-parameter family of dilaton--metric solutions of section 2 are
easily verified to form the particular case $\omega=0$ of the
class of solutions of the present section. Similarly, the limit
$\omega\to\infty$ --- which corresponds to a pure $S$--duality
transformation: $S\to-1/S$ --- also reproduces the solutions of
section 2, but with the opposite sign for $\gamma$.
All other choices of $\omega$ lead to a nonzero axion configuration,
and are therefore new solutions.

\ref\ortin{T. Ort\'{\i}n, \prd{47}{93}{3136}}

We also see in this way how the solutions $(\delta,\gamma)$ and
$(\delta,-\gamma)$ of the pure dilaton--metric
system are continuously connected by a one-parameter new class of
solutions with a nonvanishing axion, corresponding to varying
$\omega$ from zero to infinity. Notice that all of these
solutions are still singular at $r=\ell$, except for the special case
$\gamma=0$. Also notice that, even though $Q_\ssd$ vanishes
for the parameter values $\omega=\pm1$, the presence of the axion
field still induces a nonvanishing dilaton background. A final point
of interest is that, for all values of $\omega$, the combination
$Q_{\ssd}^2+Q_\ssa^2=\gamma^2\ell^2$ remains fixed
(a similar observation was made in Ref.~\ortin\ ).
We see that our \sltwor\ transformations can be characterized as a
rotation in the space of these two scalar charges --- explicitly,
if we set $\omega=(1-\sin\Theta)/\cos\Theta$, eq.~\charges\ reduces
to $Q_{\ssd}= \gamma \ell\sin\Theta$ and $Q_\ssa=\gamma \ell\cos\Theta$.

The solutions we have obtained are not quite the most general
solutions of the  axion--dilaton--metric system in four dimensions
that are static, spherically symmetric and asymptotically flat. This
is because, following ref.~\moresolns, one can add a purely topological
contribution to the antisymmetric tensor: $B_{\theta\varphi} =
Q_{\rm top}\sin\theta$. This potential is spherically symmetric since
it yields a field strength, $H=dB$, which completely vanishes. It
nevertheless cannot be gauged away provided that the
second homotopy group of the background spacetime is nontrivial
\moresolns. Even though its field strength vanishes, such a topological
configuration can have real physical effects for macroscopic strings
in both the bosonic and heterotic string theories.
Further note that one cannot introduce this topological charge
when working with the pseudoscalar representation of the axion.

We therefore arrive at a three-parameter family of metric--dilaton--axion
configurations, which precisely corresponds to the three physical quantities,
$M$, $Q_\ssd$ and $Q_\ssa$, we expect to describe the asymptotic falloff of
our three kinds of fields.  Therefore (putting aside the topological
exception just discussed) the solution we have
obtained --- {\it viz} the metric of eqs.~\ansatzt\ and \answertt, together
with
the dilaton and axion of eq.~\firstt\ --- are the most general such static,
spherically symmetric, and asymptotically flat low-energy string configuration.

\subsection{The `Taub--NUT' Case}

\ref\buscher{T. Buscher, \plb{194}{87}{59}; \plb{201}{88}{466}.}
\ref\welch{D.L. Welch, preprint UCSBTH-94-15, hep-th/9405070}
\ref\AALb{E. Alvarez, L. Alvarez-Gaum\'e and Y. Lozano, ``A Canonical
Approach to Duality Transformations'',
preprint CERN-TH-7337/94, hep-th/9406206.}

We next generate a slightly more general class of solutions, for which the
metric is not static, but is stationary. To do so we perform a target-space
duality
transformation based on the timelike isometry of time translation.
The action of such a duality transformations for a nontrivial configuration
involving the metric, dilaton and antisymmetric tensors is given by
\buscher:
\label\dualbusch
\eq\eqalign{
\tilde{G}_{tt}&=1/G_{tt},\qquad
         \tilde{G}_{ti}=-B_{ti}/G_{tt},\qquad
          \tilde{G}_{ij} = G_{ij} -
{G_{ti}G_{tj} - B_{ti} B_{tj}\over G_{tt}}\cr
\tilde{B}_{ti}& = -{G_{ti}}/{G_{tt}},\qquad
        \tilde{B}_{ij}=B_{ij}+{G_{ti}B_{tj}
         -G_{tj}B_{ti}\over G_{tt}},\qquad
e^{\tilde\phi}=e^\phi\left({\det G\over \det\tilde G}\right)^{1/2}}
\eeq
where, in our case `$t$' denotes  the time direction. As for earlier sections,
$G_{\mu\nu}=e^{-\phi} g_{\mu\nu}$ here represents the sigma-model
metric.\foot\funnysign{The sign we give here for the transformation of the
`$\ss t-i$' components of the fields is the opposite of what is often found in
the literature \buscher, but this can be corrected by performing the coordinate
transformation $\ss t\to -t$ in the dual solution.
A similar result was found in Ref.'s \welch, \AALb.
As presented, the duality transformation is closely related to the
$\ss O(d,d+p)$ transformations applied in the next section.}

Applying these transformations to our general metric--dilaton--axion
backgrounds of the previous section, eqs.~\ansatzt\ and \answertt\ for the
metric, together with the dilaton and axion configurations of eq.~\firstt,
we are led to a set of dual solutions to the low-energy string equations,
which we denote by a tilde. The dual Einstein metric is given by:
\label\dualeinstt
\eq
d\tilde{s}^2 = - e^{\hat\phi}\left(dt+Q_\ssa\cos\theta\,
 d\varphi\right)^2 +
e^{-\hat\phi}\left[\, dr^2 + \Lambda r^2
( d\theta^2 + \sin^2\!\theta \,
d\varphi^2 )\right]
\eeq
where $e^{-\hat\phi}$ represents the quantity given in eq.~\firstt,
and
\label\dualphib
\eq
e^{\tilde\phi}=f=\Lambda^\delta, \qquad \tilde B_{\mu\nu}=0.
\eeq

This new solution is reminiscent of the Taub--NUT solution \taubnut\
of the vacuum Einstein equations because of the appearance of
$(dt + Q_\ssa \cos\theta\,d\varphi )^2$ in the line element.
In fact, eqs. \dualeinstt\ and \dualphib\ are an extension of the
Taub--NUT solution to low-energy string theory which includes an
arbitrary dilaton charge. In this solution, the dilaton charge is
$\tilde Q_\ssd=\delta \, \ell$ while the mass is given by $2\GN \tilde
M = {1-\omega^2\over1+\omega^2}\,\gamma\ell$. For $\delta=0$, the
dilaton vanishes and we recover precisely the Taub--NUT metric, of
which the standard form can be achieved by a change of variables:
$\tilde r=r-{\omega^2\ell\over\omega^2+1}$.

We define the NUT parameter, $N$,  in terms of the coefficient appearing
in the time differential by, $dt+2 N \cos\theta\,d\varphi$. Thus
the NUT parameter, $\tw{N}$, of the dual solution is given in terms
of the axion charge, $Q_\ssa$, of the original configuration by
$\tilde N = Q_\ssa/2 ={\omega\gamma\ell\over1+\omega^2}$.

When $\delta$ is not zero, the surface $r=\ell$ contains a real
curvature singularity. There are also conical singularities at the axes,
$\theta = 0$ and $\theta = \pi$, unless the time coordinate happens to
be periodically identified with period $8\pi N$ \misner. This metric
\dualeinstt\ is both time-translation invariant and spherically symmetric,
but these symmetries act more subtly than they did on our previous
examples. In particular, the rotational symmetries act in the usual way
on the angular coordinates, but also involve time translations in
order to preserve the differential $ dt+2N \cos\theta\,d\varphi$.
Thus we have lost spherical symmetry, in the conventional sense.
One finds then that surfaces of constant radius have
the topology of a three-sphere, in which there is a Hopf fibration
of the $S^1$ of time over the spatial $S^2$ \misner. Note that this
interesting topology was created in the duality transformation,
by the exchange of the axion charge for the NUT parameter.

One can reobtain the general case with nonvanishing axion field
by performing a further \sltwor\ transformation on the solution, \dualphib,
just constructed. This lead us in the present case to the same Einstein metric
as in eq.~\dualeinstt, but with new dilaton and axion fields, $\phi'$ and $a'$.
These are given explicitly by:
\label\firsttt
\eq
e^{\phi'} = \left(1+\epsilon^2\right){\Lambda^\delta
\over\epsilon^2\Lambda^{2\delta}+1} \qquad \hbox{and} \qquad
a^\prime ={\epsilon\left(\Lambda^{2\delta}-1\right)\over
\epsilon^2\Lambda^{2\delta}+1} ,
\eeq
where $\epsilon$ is the parameter of the new \sltwor\  transformation.

Using eq.~\axion, we see
\label\bfirstt
\eq
{B^\prime_{\varphi t}}(\theta)=Q'_\ssa\cos\theta
\eeq
where new axion charge is $Q'_\ssa={2\epsilon\delta\ell\over 1+\epsilon^2}$.
An asymptotic expansion of the new dilaton field shows that this solution's
dilaton charge is given by $Q'_{\ssd}={1-\epsilon^2\over 1+\epsilon^2}
\, \delta\ell$. Since the Einstein metric is unchanged by the \sltwor\
transformation, so are the mass and NUT parameters, which remain
as given above.

Clearly, since repeated applications of \sltwor\ and duality transformations
have generated new classes of solutions, these two transformations
do {\it not} commute. One might wonder at this point if their repeated
application would continue to generate new classes of solutions.
Fortunately, applying \dualbusch\ to our latest class of solutions
simply gives back more solutions within the same class, and so no
further solutions are generated in this way. This is as would be expected
since the four parameters of these solutions exhaust the four
quantities which define the asymptotic falloff of the fields we consider.

The three-dimensional `moduli' space of the solutions that we have
obtained, for a fixed value of $\ell$, turns out to be compact.  This is
because ($i$) $\delta$ and $\gamma$ are restricted by
$\delta^2+\gamma^2=1$; and ($ii$) changing $\omega\rightarrow
1/\omega$ leaves the solutions invariant provided that $\gamma$
is taken to $-\gamma$ at the same time. Thus we can restrict to
values $\omega\leq 1$. Finally, ($iii$) the parameters $\epsilon$ and
$\delta$ are identified in precisely the same way as are $\omega$
and $\gamma$, and so it suffices to consider $\epsilon\leq 1$.

We next discuss the singularities in our four-parameter class of
solutions. Singularities occur at $r=0$ when $\ell<0$, and at
$r=\ell$ when $\ell>0$. The latter becomes a nonsingular surface
for $\delta=0$ and $\gamma=1$. (\cf\ eq.~\firsttt, which shows that
the parameter $\epsilon$ is irrelevant whenever $\delta=0$.)
If both $\omega = \delta =0$, then we recover
the Schwarzschild solution, for which $r=\ell>0$ is the event horizon,
and $r=0$ is a curvature singularity.
(As before, the case $\delta=0,\, \gamma=-1$ corresponds to a
negative mass Schwarzschild solution, without a horizon.)

Nonvanishing $\omega$ (with $\delta=0$) yields the
Taub--NUT solution, where again $r=\ell$ is only
a coordinate singularity. If the time were chosen to
be periodic in this solution, as discussed above,
this surface would not be a global event horizon, although
it would still be an apparent horizon. In this case, the geometry
is entirely free of singularities \misner, and one may extend the
radial coordinate to $-\infty$.

Summarizing this section, we have found the most general static
asymptotically flat, spherically symmetric background for the
axion--dilaton--metric system in four-dimensions (apart from
the possibility of a topological\foot\throwaway{We note in passing that in the
larger class of solutions, for which the NUT parameter is nonzero and the
time coordinate is periodically identified, the spacetime's second homotopy
group is trivial, and so this topological field configuration is pure gauge.}
 contribution $B_{\theta\varphi}=
Q_{\rm top}\sin\theta$). Duality and \sltwor\  transformations
naturally extend this solution to one including a non-trivial
NUT parameter as well, given by the Einstein  metric \dualeinstt,
the dilaton \firsttt\  and the antisymmetric tensor \bfirstt.
The Killing symmetries of this solution are still time translations,
and $SO(3)$ rotations. Spherical symmetry in a conventional sense
is lost, though, when the NUT parameter is nonzero, since the rotations
act on the time coordinate as well. The final solution
depends on four arbitrary parameters ($\ell, \delta, \omega$ and $\epsilon$
with $\delta^2+\gamma^2=1$), which correspond to the four `physical charges'
which define the asymptotic behaviour of the fields involved: \ie, the
mass $M$,  dilaton charge $Q_{\ssd}$, axion charge $Q_\ssa$ and Taub-NUT
parameter $N$.

\section{Gauge Field Backgrounds}

We next generalize the solutions of the previous two sections to include
a nonvanishing (abelian) gauge field configuration, such as can appear
in the heterotic string. We are led in this section to a six parameter
family of axion--dilaton--metric--electromagnetic field configurations,
for which the previous four parameters are supplemented by the
solution's electric and magnetic charges. This is a much broader
class of solutions than has been obtained previously in the literature,
which has considered either ($i$) an arbitrary mass and arbitrary
electric and magnetic charges \wilczek, \kall, or ($ii$) arbitrary mass,
dilaton charge, and one of either electric or magnetic charge \italia.

Our starting point is the metric, dilaton and antisymmetric tensor fields,
respectively given by eqs.~\dualeinstt, \firsttt, and \bfirstt\ of the previous
section. The key to generalizing these configurations to the electromagnetic
case, using only algebraic manipulations, is to use the continuous
extension of the discrete duality transformation we have been using to this
point.

\ref\oddone{S. Cecotti, S. Ferrara and L. Girardello, \npb{308}{88}{436};
\bk K. Meissner and G. Veneziano, \plb{267}{91}{33};\bk
A. Sen, \plb{274}{92}{34};\bk
M. Gasperini, J. Maharana and G. Veneziano, \plb{272}{91}{277}.}
\ref\oddhet{S. Hassan and A. Sen, \npb{375}{92}{103}.}
\ref\oddgage{A. Sen, \plb{271}{91}{295}.}

It has be shown \oddone\ that whenever the string background is
independent of $d$ of the spacetime coordinates,
there exists an $O(d,d)$ symmetry which acts in the space of solutions
of the low energy field equations. The same results were extended
to the heterotic string in ref.~\oddhet. In the heterotic case, if the
solutions are independent of $d$ of the spacetime coordinates, and also
have background gauge fields which lie in a commuting subgroup
which has $p$ $U(1)$ generators, then the low-energy string equations
admit an $O(d,d+p)$ symmetry.
Provided that the spacetime has a
Minkowski-signature metric and that time translation is one of the
symmetry directions --- certainly the case of interest here --- one can
show, for infinitesimal transformations, that the generators in an
$O(d-1,1) \times O(d+p-1,1)$ subgroup of this group actually
relate distinct solutions, while the remainder generate pure gauge
transformations \oddgage.

In the present instance, we consider only time translation symmetry
and a single $U(1)$ gauge field --- \ie, $d=p=1$. We therefore expect
to be able to generalize our existing solutions using a one-parameter
family of \ooneone\ transformations. The action of these transformations
is most easily written when the background fields are written as the
following $9 \times 9$ matrix \oddhet
\label\Mmatrix
\eq
{\cal M}=\pmatrix{\cak_-^TG^{-1}\cak_-&\cak_-^TG^{-1}\cak_+&-\cak_-^TG^{-1}A\cr
              \cak_+^TG^{-1}\cak_-&\cak_+^TG^{-1}\cak_+&-\cak_+^TG^{-1}A\cr
              -A^TG^{-1}\cak_-&-A^TG^{-1}\cak_+&A^TG^{-1}A\cr}
\eeq
where
\label\Kmatrix
\eq
(\cak_\pm)_{\mu\nu}=-B_{\mu\nu}-G_{\mu\nu}-{1\over4}A_\mu A_\nu
\pm\eta_{\mu\nu}
\eeq
and $\eta_{\mu\nu}$ is the flat Minkowski metric in four dimensions.
In order to make contact with the previous literature, we adopt here the
convention that time is the fourth component --- \eg,
$\eta_{\mu\nu}={\rm diag}(1,1,1,-1)$.

The \ooneone\ symmetry can be expressed as the invariance of
the low-energy string equations under the transformation
$\cam\to \camb=\Omega \cam\Omega^T$, where the \ooneone\
transformation matrix is given by
\label\Omatrix
\eq
\Omega=\pmatrix{I_7&0&0\cr0&x&
\sqrt{x^2-1}\cr0&\sqrt{x^2-1}&x} .
\eeq
Here $I_7$ represents the $7\times7$ unit matrix, and $x$ is
a parameter which satisfies $x^2 \geq 1$.  The dual fields,
which we denote by $\Gbar_{\mu\nu}$, $\Bbar_{\mu\nu}$ and
$\Abar_{\mu}$ can then be found by re-expressing $\camb$ in the
form \Mmatrix.  The symmetry also acts on the dilaton field according
to the rule
\label\dilatonp
\eq
e^{\phi^\prime}= \left({{\rm det}G\over{\rm
det}G^\prime}\right)^{1\over2} e^\phi\ \ .
\eeq

The \ooneone\ transformations we have just defined fall into
two disconnected classes that are characterized by the
sign of $x$, since either $x \geq 1$ or $x\leq -1$.  Ordinary discrete
duality as it has been used so far in the text, simply interchanges
these two classes. For example, $\Omega(x=1)$ is the identity
transformation, while $\Omega(x=-1)$ generates the dual background.

Actually the result of composing two transformations
$\Omega(x)\Omega(-1)$ gives a transformation in which one reverses
both the sign of $x$, and the sign of the off-diagonal terms in
eq.~\Omatrix. The effect of the latter sign change is simply
to reverse the sign of the electromagnetic fields. This leads us
to decompose $\Omega(-1)$ in terms of two commuting matrices:
$\Omega(-1) = \Omega_\ssd\Omega_q$ where $\Omega_q=
{\rm diag}(1,1,1,1,1,1,1,1,-1)$ and $\Omega_\ssd={\rm diag}
(1,1,1,1,1,1,1,-1,1)$. Applying $\Omega_q$ to transform $\cam$
changes the sign of the gauge field, while $\Omega_\ssd$
generates the duality transformation of
eq.~\dualbusch.\foot\moresigns{Notice that $\ss \Omega'_\ssd=
{\rm diag}(1,1,1,-1,1,1,1,1,1)$ generates the duality transformation
with the conventional signs \buscher.}

We now apply these transformations to the dilaton--metric--axion
configurations of the previous section. We start by converting the
Einstein metric of eq.~\dualeinstt\ to the sigma-model metric, finding:
\label\sssmetrict
\eq
dS^2 = - F(r) \,
\left(dt+Q_\omega\cos\theta\, d\varphi\right)^2 +
G(r)\, dr^2 + H^2(r) ( d\theta^2 + \sin^2 \theta \,
d\varphi^2 ) ,
\eeq
with
\label\secondd
\eq
\eqalign{
F(r)&\equiv e^{\hat\phi-\phi^\prime}=\left({1+\omega^2\over
1+\epsilon^2}\right)\, \left({\epsilon^2\Lambda^{2\delta}+1
\over \omega^2\Lambda^{2\gamma}+1}\right)\,
\Lambda^{\gamma-\delta}\cr
G(r)&\equiv
e^{-(\hat\phi+\phi^\prime)}={\left(\epsilon^2\Lambda^{2\delta}+1\right)
\left(\omega^2\Lambda^{2\gamma}+1\right)
\over \left(1+\epsilon^2\right)\left(1+\omega^2\right)}\,
\Lambda^{-\gamma-\delta}\cr
H^2(r)&\equiv r^2\,\Lambda\,e^{-(\hat\phi+\phi^\prime)} =r^2\,
\Lambda\, G(r)}.
\eeq
In these expressions $\Lambda(r)=1-\ell/r$, as in previous sections,
and $Q_\omega ={2\omega\gamma \ell\over1+\omega^2}$.
We take the dilaton and antisymmetric tensor fields from
eqs. \firsttt\ and \bfirstt, respectively (dropping the primes)
\label\bfirsttt
\eq\eqalign{
e^{\phi}&=\left(1+\epsilon^2\right){\Lambda^\delta
\over\epsilon^2\Lambda^{2\delta}+1}\cr
B_{\varphi t}&=Q_{\epsilon}\cos\theta\cr}
\eeq
where $Q_{\epsilon}={2\epsilon\delta \ell\over1+\epsilon^2}$.

Applying the \ooneone\  transformation to these solutions, we generate a
new class of solutions which depends on an additional parameter, $x$. The
sigma-model metric of this new class is given by:
\label\sssmetrictd
\eq\eqalign{
d\overline{S}^2 & = - {F(r)\over J(r)^2}
\,
d\xi^2 + G(r) \, dr^2 + H(r)^2
(d\theta^2 + \sin^2\theta \, d\varphi^2),\cr
\Bbar_{\varphi t}&=\left({(1+x)\, Q_\epsilon+ (1-x)\, Q_\omega\, F(r)\over 2\,
J(r)}\right)\, \cos\theta ,\cr
\Abar_t &=\sqrt{x^2-1}\left({1-F(r)\over J(r)}\right),\cr
\Abar_\varphi &=\sqrt{x^2-1}\left({Q_\epsilon-Q_\omega\, F(r)\over
J(r)}\right)\, \cos\theta ,\cr
 e^{\phibar}&=e^{\phi}J(r)=\left(1+\epsilon^2\right){\Lambda^\delta
\over\epsilon^2\Lambda^{2\delta}+1}\, J(r)}
\eeq
where
\label\lineelem
\eq
d\xi\equiv dt+\left({1-x\over 2}\, Q_\epsilon+{1+x\over 2}\,
Q_\omega\right)\, \cos\theta\, d\varphi\equiv dt+2\Nbar \, \cos\theta\,
d\varphi
\eeq
and
\label\jota
\eq
J(r)\equiv {1\over 2}\left[ (1+x)+(1-x)\, F(r)\right] .
\eeq
All the other components of $\Gbar_{\mu\nu}$, $\Bbar_{\mu\nu}$
and $\Abar_\mu$ turn out to vanish. Eq.~\lineelem\ defines the NUT
parameter, $\Nbar$, for this new metric.

For future use, we record here the dilaton and axion charges
for the above solution:
\label\dilchargex
\eq
\eqalign{
\Qbar_\ssd&={\ell\over 2}\left[(1+x){1-\epsilon^2\over 1+\epsilon^2}\,
\delta+(1-x){1-\omega^2\over 1+\omega^2}\, \gamma\right]\cr
\Qbar_\ssa&=\ell\left[(1+x){\epsilon\delta\over 1+\epsilon^2}+
(1-x){\omega\gamma\over 1+\omega^2}\right]\ \ .\cr}
\eeq
These are extracted from the asymptotic expansions of the dilaton
and antisymmetric tensor fields, where for the latter,
$\Bbar_{\varphi t}(r\rightarrow\infty) \to Q_\ssa \cos\theta$
as $r \to \infty$.

Next, we perform an \sltwor\  transformation, thereby introducing
another free parameter into our class of solutions. Since this symmetry
is defined to act on the dilaton and axion fields, $\phi$ and $a$, it
is first necessary to determine $a$ from the given expression for
$B_{\mu\nu}$, using eq.~\axion. This requires knowledge of the antisymmetric
field strength tensor, $H_{\mu\nu\rho}$, for which we not only need
the curl of $B_{\mu\nu}$, but also the corresponding gauge-field
Chern--Simons terms, since these no longer vanish for the configurations
we are considering.
\label\hmunu
\eq
\Hbar_{\mu\nu\rho}=\partial_\mu \Bbar_{\nu\rho}-{1\over 4}\, \Abar_\mu\,
\Fbar_{\nu\rho}
+ {\rm cyclic\, \,  permutations}
\eeq

Now, inspection of eqs.~\sssmetrictd\ shows that the nonvanishing
components of the gauge field strength are $F_{tr}$, $F_{r\varphi}$
and $F_{\theta\varphi}$. Using these, as well as the expression for
$\Bbar_{\mu\nu}$ from eq.~\sssmetrictd, in eq.~\hmunu, we see
that $\Hbar_{rt\varphi}$ vanishes even though $\Bbar_{\varphi t}$
is a function of $r$. This provides a nontrivial check of our results,
since a nonvanishing $\Hbar_{rt\varphi}$ would have
implied a $\theta$ dependence for the scalar axion field,
in contrast with the requirements of the $SO(3)$ rotational
symmetry. The only nonvanishing component of
$\Hbar_{\mu\nu\lambda}$ turns out to be $\Hbar_{\theta\varphi t}$,
from which we obtain
\label\axiondd
\eq
\abar(r)=\left({1-x\over 2}\right)\,
{\omega\left(\Lambda^{2\gamma}-1\right)\over
\omega^2\Lambda^{2\gamma}+1}+\left({1+x\over
2}\right){\epsilon\left(\Lambda^{2\delta}-1\right)\over
\epsilon^2\Lambda^{2\delta}+1}.
\eeq
We choose here an arbitrary integration constant to ensure that
$\abar(r)$ vanishes at infinity.

We now wish to perform the \sltwor\ transformation to these
configurations. In this case, since the gauge field background
does not vanish, we must use a more general transformation rule.
Not only must the dilaton and axion fields in $\ol{S}=\abar+
i\, e^{\phibar}$ be transformed as in eq.~\sltr, but we must also
transform the gauge fields \wilczek, \slur. The total transformation
becomes
\label\sltrt
\eq
\eqalign{
S&\longrightarrow {{\rm a}\, S + {\rm b}\over {\rm c}\,
S +{\rm d}}\cr
(F_+)_{\mu\nu}&\longrightarrow \left({\rm c}\, S +{\rm d}\right)\,
\left(F_+\right)_{\mu\nu}\cr
(F_-)_{\mu\nu}&\longrightarrow \left({\rm c}\,  S^* +{\rm d}\right)\,
\left(F_-\right)_{\mu\nu}}
\eeq
where $(F_\pm)_{\mu\nu} \equiv F_{\mu\nu}\pm {i\over2} \;
\epsilon_{\mu\nu\rho\kappa} \, F^{\rho\kappa}$ are respectively
the (Hodge) self-dual and the antiself-dual parts of the electromagnetic field
strength, $S^*$ is the complex conjugate of $S$.
Again, the Einstein metric, which is required to define
the volume form and the contractions in $(F_\pm)_{\mu\nu}$,
is left invariant under this transformation. As in the previous section,
all but one of the three \sltwor\ parameters are eliminated by the requirement
that the dilaton and axion fields must vanish at infinity. We denote
the single extra parameter which remains by $\rho\equiv {\rm c}/{\rm d}$.

Applying \sltwor\  to the solutions \sssmetrictd, we obtain
the new dilaton and axion fields, $\hat{\phi}$ and $\hat{a}$:
\label\thirddil
\eq
\eqalign{
e^{\hat\phi}&={\left(\rho^2+1\right)\, e^{\phibar} \over
\rho^2\, e^{2\phibar}+\left(\rho\, \abar+1\right)^2}\cr
\hat{a}&={\rho\left(\abar^2+e^{2\phibar}-1\right)-
\left(\rho^2-1\right)\, \abar\over \rho^2\, e^{2\phibar}+\left(\rho\,
\abar+1\right)^2}}
\eeq
{}From which we find the corresponding charges
\label\thirdcharge
\eq\eqalign{
 {\widehat Q}_{\ssd}&={1-\rho^2\over 1+\rho^2}\, \Qbar_{\ssd}-{2\, \rho
\over 1+\rho^2}\, \Qbar_\ssa\cr
\widehat Q_\ssa&={1-\rho^2\over 1+\rho^2}\, \Qbar_\ssa+{2\, \rho
\over 1+\rho^2}\, \Qbar_\ssd}
\eeq
where $\Qbar_\ssd$ and $\Qbar_\ssa$ are given in  eq.~\dilchargex.
Notice that for $\rho\rightarrow 0$ both reduce to their previous values,
and that again the \sltwor\ transformation acts here to
rotate the charges preserving: $\Qbar_\ssd^2+\Qbar_\ssa^2=
{\widehat Q}_\ssd^2+{\widehat Q}_\ssa^2$.

For the gauge fields, eq.~\sltrt\ also implies the following new field
strength tensor:
\label\gaugefthree
\eq
\widehat F_{\mu\nu}={1\over\sqrt{1+\rho^2}}\left[\left(1+\rho\,\abar\right)\,
\Fbar_{\mu\nu}-{1\over 2}\rho e^{\phibar} \epsilon_{\mu\nu\rho\sigma}
\, \Fbar^{\rho\sigma}\right] .
\eeq
Once again, the only nonvanishing components are
$\widehat F_{\theta\varphi}, \widehat F_{tr}$ and $\widehat F_{r\varphi}$.
A gauge potential which produces this field strength, is given by:
\label\gaugepot
\eq
\widehat A_\varphi =\Psi(r)\, \cos\theta \qquad
\widehat A_t = {\left(\Psi(r)+\widehat Q_\ssm \right)\over2\Nbar} ,
\eeq
where
\label\magic
\eq
\Psi(r)=\sqrt{{x^2-1\over \rho^2+1}}\, \left[\left(1+\rho\abar\right)\,
{Q_\epsilon-Q_\omega\, F\over J}+\rho \ell\left(
{\omega^2\Lambda^{2\gamma}-1\over \omega^2\Lambda^{2\gamma}+1}\,\gamma
-{\epsilon^2\Lambda^{2\delta}-1\over \epsilon^2\Lambda^{2\delta}+1}\,\delta
\right)\right]
\eeq
and $\Nbar$ is the NUT parameter defined in eq.~\lineelem.

This electromagnetic field configuration has the following magnetic
charge
\label\magcharge
\eq
\widehat Q_\ssm =\sqrt{{x^2-1\over \rho^2+1}}\,\ell\, \left[
{\rho(1-\omega^2)+2\omega\over1+\omega^2} \; \gamma
-{\rho(1-\epsilon^2)+2\epsilon\over1+\epsilon^2} \; \delta\right] ,
\eeq
which can be determined by comparing to an asymptotic behavior of
the form: $\widehat F_{\theta\varphi}\simeq\widehat
Q_{\ssm}\, \sin\theta$, or $\widehat A_\varphi\simeq-\widehat
Q_{\ssm}\, \cos\theta$. In eq.~\gaugepot, we have used an arbitrary
constant that appears in solving for $\widehat A_t$ by requiring that
$\widehat A_t$ vanish as $r\rightarrow\infty$.

The electric charge of this configuration is similarly given by:
\label\elecharge
\eq
\widehat Q_\sse =\sqrt{{x^2-1\over \rho^2+1}}\,\ell\, \left[
{1-\omega^2-2\rho\omega\over1+\omega^2} \; \gamma
-{1-\epsilon^2-2\rho\epsilon\over1+\epsilon^2}\; \delta\right]
\eeq
as may be determined from the asymptotic behaviour: $\widehat
F_{tr}\simeq\widehat Q_\sse/r^2$ or $\widehat A_t\simeq
\widehat Q_\sse/r$.
One may verify that as expected the \sltwor\ transformation
rotates the electric and magnetic charges amongst each other,
preserving ${\widehat Q_\sse}^2+{\widehat Q_{\ssm}}^2$.

Finally, we can determine the antisymmetric tensor field,
$\widehat B_{\mu\nu}$ from our expression for $\hat a$ and
$\hat A_\mu$, by using eqs.~\axion\ and \hmunu. The only
component which can be nonvanishing is
$\widehat B_{\varphi t}$, and this is given by
\label\guess
\eq
\widehat B_{\varphi t}= \left[ {\widehat Q_\ssm( \Psi(r)+\widehat Q_\ssm)
\over8\Nbar}+\widehat Q_\ssa\right] \, \cos\theta\ \ .
\eeq
Notice that, asymptotically, $B_{\varphi t}\to \widehat Q_\ssa\, \cos\theta$
as expected. It also reduces to its previous expression in the limit
$\rho\to 0$.

Since the Einstein metric is left untouched by \sltwor\ transformations, it
can be read directly from eq.~\sssmetrictd:
\label\metricthree
\eq
{d\hat s}^2  = e^{{\phibar}} \, \left( - {F(r)\over J(r)^2} \,
(dt+2\Nbar \, \cos\theta\, d\varphi )^2 + G(r)
\, dr^2 + H(r)^2
(d\theta^2 + \sin^2\theta \, d\varphi^2)\right) ,
\eeq
and the sigma-model metric is obtained from this by using
$\widehat G_{\mu\nu}=e^{-\hat\phi}\gbar_{\mu\nu}$.
As was determined earlier, \cf\ eq.~\lineelem, the Taub--NUT
parameter for this solution is $\Nbar={1-x\over4}\ Q_\epsilon+
{1+x\over4}\ Q_\omega$.  Finally, an asymptotic expansion of
the Einstein metric gives the masses of these solutions to be
\label\massesx
\eq
{\widehat M} =
{\ell\over 4 G_N}\left[(x+1){1-\omega^2\over
1+\omega^2} \;  \gamma-(x-1){1-\epsilon^2\over 1+\epsilon^2}\; \delta\right]
\ \ .
\eeq
Notice that these metrics have new singularities, that are
associated with the radius $r=r_s$, for which $J(r_s)=0$, in addition to
the previous ones that are located at $r=0$ and $r=\ell$.

\section{Conclusions}

We have finally arrived at a six-parameter family of
backgrounds --- the six parameters being
$\ell, \delta, \omega, \epsilon, x$ and $\rho$.
These six parameters can be traded for six physical constants
which characterize the asymptotic form of the solutions:
the mass,  Taub-NUT parameter, axion charge, dilaton charge,
and electric and magnetic charges.
Within this family we find the most general class of
four-dimensional solutions of the leading order string field
equations, which are spherically symmetric, static and
asymptotically flat. This subclass is obtained by setting the
Taub-NUT parameter to zero.
The figure gives an overview of our construction, and
shows the origin of each of the independent parameters.

Many of these solutions have singularities, and for this reason
they have not appeared in many discussions of string corrections to
black hole spacetimes. We have kept them here since
we regard it as an open question whether the nominally
singular solutions provide legitimate backgrounds for
nonsingular string propagation.
In any event, in many cases (such as  at $r=\ell$
for the $d=4$ dilaton--metric solution, with $0 < \delta < 1$)
the singularity is
lightlike, and so no asymptotic observer can see them. Thus they
need not be regarded as `naked' in the strictest
sense. The same is not true when $-1 < \delta
\le 0$, since in this case the singularity at $r=\ell$ is timelike
and without a horizon.

\ref\will{For the classic review see, C.M. Will, {\it Theory and Experiment
in Gravitation Physics} (revised edition), Cambridge University Press, 1993.}

If it should turn out that the dilaton were to be light enough to be of
interest for systems of astrophysical size, then the existence of this
new class of scalar-metric solutions  to the low-energy string
equations could have interesting consequences.
This is because they do not appear to have been included amongst the
alternatives to general relativity that are traditionally considered
when theories of gravitation are confronted with experiment \will.
We hope to generalize the usual treatment
in a future publication.

\ref\duff{M. Duff, CERN seminar, September 1994;
M.J. Duff, S. Ferrara, R.R. Khuri and J. Rahmfeld,
in preparation; \bk
A. Sen, ``Strong-Weak Coupling Duality in Three-Dimensional
String Theory'', preprint TIFR-TH-94-19 (hep-th/9408083).}

These solutions were obtained by performing a succession of
\sltwor\ and \ooneone\ symmetry transformations of the
low-energy string equations,
starting from the much simpler two-parameter family of
dilaton--metric solutions of section 2. This type of algebraic
manipulation is much easier to perform than would be a
direct assault on the integration of the corresponding string
field equations. Notice that it was the failure of the
\sltwor\ and  \ooneone\  transformations to commute with
one another  which allowed us to build up the full six-parameter
family from the original two-parameter class of
solutions. Note that this is generically true for $O(d,d+p)$
and \sltwor transformations, and in fact both sets of transformations
are subsumed within a larger group of transformations \duff.

We claim that further \sltwor\  transformations
or \ooneone\  boosts (in the time-gauge directions as in eq.~\Omatrix)
will not introduce any new solutions, but only map these solutions
amongst themselves. Our reason for making this assertion
is that since we have generated all of the possible charges
which can describe the asymptotic behaviour of our fields
at infinity, we have exhausted the space of solutions to
the low-energy string equations which satisfy our stated
symmetry ansatz for the fields.  But since all of the
\sltwor\ and \ooneone\ transformations preserve these
symmetries, as well as the asymptotic flatness, they must
lead to configurations that lie within the existing class.

It is instructive to consider some of the limits of our most general
solution.
\item{1.}
Setting $x=\pm 1$ leads to vanishing gauge fields, and
gives back the general dilaton--axion--metric solutions of section 3.
Notice that $\rho$ becomes a redundant parameter when $x= \pm 1$.
\ref\robtnut{C.V. Johnson and R.C. Myers, preprint IASSNS-HEP-94/50,
McGill/94-28, hep-th/9406069 (1994); IASSNS-HEP-94/65,
McGill/94-40, hep-th/9409177 (1994).}
\ref\kallosh{R. Kallosh, D. Kastor, T. Ort\'{\i}n and T. Torma,
 preprint SU-ITP-94-12, hep-th/9406059\bk
C.M. Hull (unpublished);\bk D.V. Gal'tsov and O.V. Kechkin,
preprint hep-th/9407155.}
\item{2.}
If we instead choose $\delta=0$ (and $\gamma=1$), then
the parameter $\epsilon$ vanishes from the solutions. In this
limit we recover the Taub--NUT dyons recently constructed
in Ref.'s \robtnut\ and \kallosh.
\item{3.}
If, in addition to $\delta = 0$ and $\gamma = 1$,  we also set
$\omega=0$, then we obtain the electrically and magnetically
charged dilatonic black holes obtained in Ref.~\wilczek, \kall.
\item{4.}
Finally, the magnetically charged dilatonic black holes
of Ref.~\gibbons\ can be produced with the limit $\rho\to\infty$.
Alternatively, $\rho=0$ yields black holes with a purely electric
charge.

\ref\kot{R. Kallosh, A. Linde, T. Ort\'{\i}n, A. Peet and
A. Van Proeyen \prd{46}{92}{5278}; E. Bergshoeff, R. Kallosh
and T. Ort\'{\i}n, ``Duality versus Supersymmetry and
Compactification,'' preprint SU-ITP-94-19, hep-th/9410230;
T. Ort\'{\i}n, ``Duality versus Supersymmetry,'' preprint
QMW-PH-94-35, hep-th/9410069.}

\ref\bak{I. Bakas, ``Space time interpretation of $S$--duality and
supersymmetry violations of $T$--duality,'' preprint CERN--TH.7473/94,
hep-th/9410104.}

We expect the techniques we have used to also have useful applications
for the construction of more complicated string solutions, as
for instance using the $\varphi$--independence of these backgrounds
to generate more general axially symmetric solutions. Also
an analysis of the supersymmetric nature of our solutions on the
lines of Ref.~\kot, would be very interesting.
Recently, an investigation of the non-commuting character of
discrete $S$-- and $T$--duality transformations
has also appeared in Ref.~\bak.

\bigskip
\centerline{\bf Acknowledgments}
\bigskip

C.B. would like to thank the Institut de Physique at the Universit\'e
de Neuch\^atel for its kind hospitality while this work was in progress.
We acknowledge useful conversations with E. Alvarez, Y. Lozano,
P. Page, J. Russo, and T. Ort\'{\i}n. This research was
partially funded by
the N.S.E.R.C.\ of Canada, les Fonds F.C.A.R.\ du Qu\'ebec, and
the Swiss National Foundation.

\bigskip

\appendix{A}{Higher-Dimensional Solutions}

We begin by generalizing the dilaton--metric solutions of
section 2 to spacetime dimensions $d \ge 4$. Following Callan
\etal\ \strbhone, we write our static and spherically symmetric
metric (in the Einstein frame) in a slightly different way
\label\ansatzz
\eq
ds^2 = - U^2 dt^2 + V^2 (dr^2 + r^2 d\Omega_{d-2})
\eeq
where $d\Omega_{d-2}$ is the standard line element on a unit $(d-2)$-sphere.
The dilaton is chosen as above, $\phi=\phi(r)$. Again allowing for arbitrary
dilaton charge, the solutions may be written as
\label\solutionn
\eq \eqalign{
U^2 &= \left( {\beta\over\alpha} \right)^{2H} \cr
V^2 &= (\alpha\beta)^{2 /( d-3)} \left( {\alpha\over\beta}
\right)^{2H/(d-3)} \cr
e^\phi &= \left( {\beta\over\alpha} \right)^{K}, \cr}
\eeq
where
\label\moree
\eq
\alpha = 1+\left({\ell\over 4r}\right)^{d-3}
\qquad\qquad
\beta = 1-\left({\ell\over 4r}\right)^{d-3}.
\eeq
The constants
$H$ and $K$ are restricted to satisfy $H^2+K^2 (d-3) / (d-2)^2 =1$.
$(H,K)=(1,0)$ yields the standard Schwarzschild geometry in isotropic
coordinates (as does $(-1,0)$, although with a negative mass).
The coefficients in eq.~\moree\ have been chosen
so that with $d=4$, the constant $\ell$
coincides with the same physical length appearing in eq.~\answer --- \ie, for
the Schwarzschild solution, $\ell=2\GN M$ but note that in
the present coordinates the horizon occurs at $r=4\ell$.
These solutions have a nonvanishing dilaton charge for
$K \ne 0$ and, in this case, $r=4\ell$ is a curvature singularity
as is easily verified by evaluating the Ricci scalar using the
equations of motion.

It is straightforward to apply an $O(1,1)$ transformation as in
section 4, to generate a gauge field for these solutions.
The final result is an Einstein metric of the form
\label\newcharo
\eq
d\hat{s}^2 = - {U^2\over W^{2(d-3)/(d-2)}} \; dt^2
+ W^{2/(d-2)}V^2 (dr^2 + r^2 d\Omega_{d-2})
\eeq
where we have applied eq.~\esmrelation, and
\label\newcharot
\eq
W={1\over2} \; \left(1+x+(1-x)U^2e^{-2\phi/(d-2)} \right)\ \ .
\eeq
The final field configuration also includes the following
dilaton and gauge fields
\label\newchart
\eq
\eqalign{
e^{\hat{\phi}}&=W\,e^\phi\cr
\widehat{A}_t&=\sqrt{x^2-1} \; \left( {1-U^2e^{-2\phi/(d-2)}\over W}
\right) \ \ .\cr}
\eeq

\ref\strbhtwo{R.C. Myers and M.J. Perry, {\it Annals of Physics}
{\bf 172} (1986), 304.}

Now asymptotically, at large radius, one has:
\label\asymptopia
\eq \eqalign{
\hat{g}_{tt} &\simeq -1+\left[4H\left(1+(x-1){d-3\over d-2}\right)
-4K(x-1){d-3\over(d-2)^2}\right]
 \left({\ell\over 4r}\right)^{d-3} + \cdots \cr
\hat{g}_{ij} &\simeq \delta_{ij} \left(1+\left[4H
\left({1\over d-3}+{x-1\over d-2}\right)-4K{x-1\over(d-2)^2}\right]
\left({\ell\over 4r}
\right)^{d-3}\right) +\cdots \cr
e^{\hat{\phi}} &\simeq 1+\left[2H(x-1)-2K
\left(1+{x-1\over d-2}\right)\right]
 \left({\ell\over 4r} \right)^{d-3} + \cdots \cr
\widehat{A}_t&\simeq\sqrt{x^2-1}\left(4H-{4K\over d-2}\right)
\left({\ell\over 4r} \right)^{d-3} + \cdots \ .\cr}
\eeq
As was the case in four dimensions, the asymptotic form of $g_{ij}$ can be
used to define the masses for these solutions. We find it to be given
by \strbhtwo:
\label\gimasses
\eq
2\GN M = {A_{d-2}  \over2 \pi }
\left[H\left(d-2+(x-1)(d-3)\right)
-K(x-1){d-3\over(d-2)}\right]\left(\ell\over 4\right)^{d-3},
\eeq
where $A_{d-2}$ is the area of the unit $(d-2)$-sphere.
Defining the dilaton charge of these solutions as the coefficient of
$-(1/r)^{d-3}$ in the asymptotic expansion of $e^\phi$, we similarly obtain
\label\hiddilcharge
\eq
Q_\ssd = \left[2 K\left(1+{x-1\over d-2}\right)
-2H(x-1)\right] \left( {\ell \over 4} \right)^{d-3}.
\eeq
Finally, using $\widehat{F}_{tr}\simeq \widehat{Q}_\sse/r^{d-2}$,
we find the electric charge to be
\label\eech
\eq
Q_\sse=\sqrt{x^2-1}\ (d-3)\left(4H-{4K\over d-2}\right)
\left({\ell\over 4} \right)^{d-3}
\eeq

\ref\hcharge{G.T. Horowitz and A. Strominger, \npb{360}{91}{197}.}

For $d>5$ dimensions, these three physical charges
completely characterize the solutions which are restricted
to be static, asymptotically flat, and spherically symmetric
(\ie, $SO(d-2)$ invariant).
For higher dimensions, there can be no magnetic charge associated
with the gauge field, nor are there any spherically
symmetric configurations of the Kalb-Ramond field. The one
exception to the latter statement is for $d=5$. In that case,
solutions may be found with an extra magnetic-like charge from
the $H$ field \gibbons\hcharge. Presumably the known solutions,
which have arbitrary masses and (magnetic) axion charges, could be
generalized to a four parameter family of solutions including
arbitrary electric gauge charges and  dilaton charges, as well.

\listrefs

\figurecaptions

\bye